\documentstyle{article}

\raggedright

 {}{\setlength\labelwidth{1.4em}\leftmargin\labelwidth
 \setlength\parsep{0pt}\setlength\itemsep{10pt}
 \setlength{\itemindent}{-\leftmargin}
 \usecounter{enumi}
 
 \sloppy
 \sfcode`\.=1000\relax}

\def\@cite#1#2{{#1\if@tempswa , #2\fi}}
\def\@biblabel#1{}
\def\x{$\pm$} 
\def\ha{H$_\alpha$~}
\def\hb{\relax \ifmmode {\rm H}\beta\else H$\beta$\fi}
\def\hi{\relax \ifmmode {\rm H\,{\sc i}}\else H\,{\sc i}\fi}
\def\hii{\relax \ifmmode {\rm H\,{\sc ii}}\else H\,{\sc ii}\fi}
\def\h2{\relax \ifmmode {\rm H}_2\else H$_2$\fi}

\def\fdg{\hbox{$.\!\!^\circ$}}
\def\farcm{\hbox{$.\mkern-4mu^\prime$}}
\def\farcs{\hbox{$.\!\!^{\prime\prime}$}}
\newcommand {\tef} {T$_{\rm eff}$}
\def\degd#1.#2{ #1\fdg#2 }                 
\def\mind#1.#2{ #1\farcm#2 }               
\def\secd#1.#2{ #1\farcs#2 }               
\def\kms{\hbox{km~s$^{-1}$}}

\begin{document} \headheight 4cm 
\thispagestyle{empty}
\begin{center}

{\LARGE SPECTROSCOPY OF NEW SUBSTELLAR CANDIDATES IN THE PLEIADES: 
 TOWARDS A SPECTRAL SEQUENCE FOR YOUNG BROWN DWARFS 
\footnote{Based on 
observations made with the William Herschel Telescope, operated on the island of La Palma by the Royal Greenwich Observatory in the Spanish Observatorio del Roque de los Muchachos of the Instituto de Astrof\'\i sica de Canarias, and on 
data collected at the 3.5m telescope of Calar Alto German-Spanish Observatory.}}

\vspace{1.5cm}

{\large }
Eduardo L. Mart\'{\i}n, Rafael Rebolo and Mar\'\i a Rosa Zapatero-Osorio \\
\smallskip
Instituto de Astrof\'\i sica de Canarias, 
E-38200 La Laguna, Tenerife, Spain \\
\bigskip
e-mail addresses: ege@iac.es, rrl@iac.es, and mosorio@iac.es
\par\vskip 0.5cm
              
\bigskip
\bigskip
\bigskip
Send offprint requests to: Eduardo L. Mart\'{\i}n 
\end{center}

\newpage

\headheight 45pt \parskip=3mm

\centerline {\large ABSTRACT}

We present optical and near-infrared spectroscopy (600--1000 nm)  
of eight faint (I$>$18) very red (R--I$>$2.2) objects discovered in a deep CCD 
survey of the Pleiades cluster (Zapatero-Osorio et al. 1996).  
We compare them with reliable cluster members like 
PPl~15 and Teide~1, and with several field very 
late-type dwarfs (M4--M9.5), which were observed with similar instrumental 
configurations. 

Using pseudocontinuum ratios we classify the new substellar candidates 
in a spectral sequence defined with reference to  
field stars of known spectral types. We also reclassify PPl~15 and 
Teide~1 in a self-consistent way. 
The likelihood of membership for the new candidates is 
assesed via the study of their photospheric features, 
\ha emission, radial velocity, and consistency of 
their spectral types and I-band magnitudes with known cluster members. 
Four of the new substellar 
candidates are as late or later than PPl~15 (M6.5), but 
only one, namely Calar~3 (M8), clearly meets all our membership 
criteria. It is indeed an object very similar to the 
brown dwarf Teide 1. 

Out of the eight new substellar candidates, our ``cautious" membership analysis 
leaves only Calar~3 as a Pleiades brown dwarf with a high level of 
confidence. This object, together with Teide~1, 
allows one to compare the 
spectroscopic characteristics of Pleiades brown dwarfs with those of old  
very cool dwarfs. The overall spectral properties are similar, but 
there are slight differences in the NaI doublet 
(818.3~nm, 819.5~nm), VO molecular band (740~nm), and some spectral ratios,   
which are probably related to lower surface gravity in the young Pleiades 
brown dwarfs than in field stars. Finally, we propose a way of improving 
future CCD-based brown dwarf surveys by using narrow-band near-IR 
pseudocontinuum filters. 

\par\vskip 1cm
{\em Subject headings:} stars: low mass, brown dwarfs, spectroscopy --
 clusters: Pleiades

\newpage 

\par\vskip 2cm \centerline{\large 1.~INTRODUCTION}

The nearby ($\sim$ 125 pc) Pleiades open cluster is now widely recognized 
as one of the best places in the Galaxy to identify free-floating brown 
dwarfs (BDs). Its  young age (70--130 Myr) implies that BDs 
should be caught at a relatively hot and luminous stage of their lives. 
Several groups 
have recognized the advantages of BD searches in this cluster, and 
consequently a number of photometric surveys have been reported 
(see Jameson 1995 for a review). 
The masses of photometric candidates can in principle 
be estimated from their position in the H-R diagram and comparison with 
theoretical evolutionary tracks. However, uncertainties 
in the conversion of photometric data to stellar parameters (\tef, Luminosity), 
and in the accuracy of model predictions have 
for many years prevented reaching a definitive conclusion 
regarding the substellar status of 
the best BD candidates found in the different surveys. 
The presumption of substellar mass for free-floating objects 
cannot be tested dynamically. Nevertheless, it  can be inferred from the  spectroscopic test, sometimes called the Li-test, proposed 
by Rebolo, Mart\'\i n \& Magazz\`u (1992). The first applications of 
the Li-test to BD candidates in binary systems, the field and in 
open clusters gave only Li non-detections (see Rebolo, Magazz\`u \& 
Mart\'\i n 1995) for a review. But very recently, the first positive 
detections have been announced in the Pleiades objects PPl~15 
(Basri, Marcy \& Graham 1996) and Teide 1 (Rebolo et al. 1996). 
The presence of Li in these two faint (I$\ge$17.8) Pleiades 
members, and its absence in slightly more luminous stars implies 
that the stellar-substellar borderline in the cluster has been crossed.

Eight new Pleiades BD candidates with I$>$18 have been identified as very red 
(R--I$>$2.2) objects in deep CCD images obtained by  
Zapatero-Osorio, Rebolo \& Mart\'\i n (1996). These objects will 
be referred throughout this paper with the abridged names given by  
Zapatero-Osorio et al, which are based on their I-band apparent magnitudes 
and the observatories where they were first observed. 
Follow-up spectroscopic observations are essential in order to 
assess the cluster membership and improve our knowledge of the atmospheres 
of Pleiades brown dwarfs. 
In this paper we present optical and near-infrared 
spectra of all the BD candidates discovered by 
Zapatero-Osorio et al. (1996). 
We have obtained  spectral types, luminosity classes, 
radial velocities and \ha emission equivalent widths. Based on all such  
information we conclude that one of the Calar objects is very similar to 
the brown dwarf Teide~1. 
We use our enlarged battery of standards for revising the spectral type 
determination of PPl~15 and Teide~1, and for discussing the differences 
between the spectroscopic properties of young brown dwarfs and main sequence 
stars.

\bigskip 

\centerline{\large 2.~OBSERVATIONS}

The observations for this programme were carried out 
on the 4.2~m William Herschel Telescope (WHT), located at the 
Observatory del Roque de los Muchachos, 
and on the 3.5~m telescope of Calar Alto Observatory. 
The observing log is shown in Table~1, and we have added 
observations of PPl~1, Teide~1 and LHS~2065 obtained in previous runs  
(Mart\'\i n 1993; Rebolo, Zapatero-Osorio \& Mart\'\i n 1995) 
with similar instrumental setups. 
The targets are 
grouped in two groups (Pleiades and field) and ranged  alphabetically. 
The apparent I-magnitudes come from Bessell (1991) and Leggett (1992) for 
the field stars, and Stauffer, Hamilton \& Probst (1994), and 
 Zapatero-Osorio et al. (1996) for the Pleiades objects. 
All of them are in the Cousins-Kron photometric system. 
  The instrumentation used was the ISIS 
double arm spectrograph at the WHT, and the twin spectrograph at Calar Alto. 
In both telescopes we removed the dichroic and used the red arm only. 
We kept the same configuration throughout the runs, namely: 
grating R158R + TEK (1124x1124) chip (WHT) and grating No. 11 + TEK 
(1124x1124) chip (3.5~m Calar).  
The spectral resolution and  coverage 
obtained with each instrument were 
5.8~\AA ~(R$\sim$1200)  and 612.6--907.0 nm; 7.8~\AA ~(R$\sim$950)  and 635.0--1025.0~nm, respectively.  

The weather conditions in both (WHT and Calar Alto) runs  
 were severely hampered 
by clouds. Sometimes a long exposure had to be paused or cut off because the 
star used for autoguiding  became invisible. Seeing was highly variable, 
ranging from 4 arcsec to less than 1 arcsec. 
In the last column of Table~1 we have marked those observations 
where we noticed passing clouds.

\begin {table} 
\caption{Log of spectroscopic observations} 
\begin{flushleft}
\begin {tabular}{lcccccc}
\hline Name  & I  & Tel. & t$_{\rm exp}$  & No. exp & Date & Weather \\
\hline 
Calar 1       & 18.2 & 3.5~m & 5400 & 2 & Oct 27, 1995 & clear \\
               &      & WHT   & 9400 & 3 & Oct 29, 1995 & clear \\
Calar 2       & 18.7 & 3.5~m & 7200 & 3 & Oct 27-29, 1995 & cloudy \\
Calar 3       & 18.7 & 3.5~m & 16200 & 5 & Oct 28-29, 1995 & clear \\
               &      & WHT   & 2320  & 1 & Oct 30, 1995    & cloudy \\
Calar 4        & 18.9 & 3.5~m  & 5400 & 2 & Oct 28, 1995 & clear \\
               &      & WHT   & 5450 & 2 & Oct 29, 1995 & cloudy \\
Calar 5       & 19.0 & 3.5~m  & 10200 & 4 & Oct 28-29, 1995 & cloudy \\
Calar 6      & 19.2 & WHT   & 6800 & 2 & Oct 30, 1995 & clear \\
Calar 7       & 19.7 & WHT   & 6800 &  2 & Oct 30, 1995 & clear \\
PPl 1         & 17.5 & INT   & 3600 &  2 &  Jan 23, 1992 & clear \\
PPl 14         & 17.4 & WHT   & 1800 &  1 & Oct 29, 1995 & clear \\
PPl 15         & 17.8 & WHT   & 3600 &  3 & Oct 29, 1995 & clear \\
Roque 1        & 18.4 & WHT   & 9000 &  3 & Feb 17, 1996 & clear \\
Teide 1        & 18.8 & WHT   & 16600 &  7 & Dec 29-30, 1994 & clear \\
BRI 0021-0214  & 15.1 & 3.5~m  & 1200 &  1 & Oct 27, 1995 & clear \\
               &      & WHT   & 4000 &  3 & Oct 28-29, 1995 & cloudy \\
GJ 51          & 10.4 & 3.5~m  & 30 &  1 & Oct 29, 1995 & clear \\
               &      & WHT   & 480 & 1 & Oct 28, 1995 & cloudy \\
GJ 65 AB       & 8.3  & WHT   & 400 & 1 & Oct 28, 1995 & cloudy \\
GJ 83.1        & 9.2  & WHT   & 300 & 1 & Oct 28, 1995 & cloudy \\
GJ 402         & 8.9  & 3.5~m  & 30  & 1 & Oct 30, 1995 & cloudy \\
GJ 873         & 7.6  & 3.5~m  & 25 &  2 & Oct 29, 1995 & clear \\
GJ 905         & 8.8  & 3.5~m  & 80 &  2 & Oct 29, 1995 & clear \\
LHS 248        & 10.5 & 3.5~m  & 50 &  2 & Oct 30, 1995 & cloudy \\
               &      & WHT & 400 &  2 & Oct 30, 1995 & clear \\
LHS 2065       & 14.5 & WHT & 600 & 1 & Dec 30, 1994 & clear \\
LP 412-31      & 15.0 & 3.5~m  & 1000 & 1 & Oct 28, 1995 & clear \\
PC 0025+0447   & 18.5 & 3.5~m  & 5400 &  2 & Oct 29, 1995 & clear \\
VB10           & 12.8 & 3.5~m  & 1200 & 1 & Oct 28, 1995 & cloudy \\   
               &       & WHT   &  900 & 2 & Oct 28, 1995 & clear \\
\hline
\end{tabular}
\end{flushleft} 
\vskip 2mm
Notes: GJ 65 A and B were not resolved on the slit due to poor seeing. 
The acronyms for the telescopes stand for 2.5~m Isaac Newton telescope 
(INT), 4.2~m William Herschel telescope (WHT) and 3.5~m Calar Alto telescope. 
\end{table}

Each individual spectrum was reduced by a standard procedure using 
IRAF\footnote{IRAF is distributed by the National Optical Observatory, which is 
operated by the Association of Universities for Research in Astronomy, Inc., 
under contract with the National Science Foundation.}, which included 
debias, flat field, optimal extraction and wavelength calibration 
using arc lamps. The rms of the wavelength calibration was always better 
that 1/10 of a pixel. 
Finally, the spectra were  
flux calibrated using the standard HD~19445, 
which has absolute flux data available in the IRAF environment.   
In Figure~1 we present the final spectra of BD 
candidates observed with the WHT, together with PPl~14, PPl~15, Teide 1 and 
three M5--M9 V standards. 
The spectra were ordered by increasingly late spectral type.    
In Figure~2 we show spectra obtained with the 3.5~m Calar Alto telescope. 
The Isaac Newton telescope spectrum of PPl~1 has a resolution of 5.8 \AA ,
similar to the WHT spectra, but it is quite noisy. 

\bigskip 

\centerline{\large 3.~ANALYSIS}

The main photospheric features of very late type dwarfs that can be 
studied with 
low resolution optical-NIR spectra are the molecular bands of CaH, TiO and VO,  
and the strong atomic lines of NaI and KI.  In principle they can 
provide us information about basic atmospheric properties like 
luminosity, metallicity and temperature. We expect that  
cluster stars share the same age and metallicity, and thus they provide 
a homogeneus sample for learning about how the atmospheres change towards  
 lower masses. We can also learn about the problem of chromospheres  
by measuring the strength of \ha in emission.  In this section 
we start presenting our method for quantitatively measuring spectral types, 
which is based on flux ratios. There is a good correlation between 
spectral type (in the range M4--M9.5) and some (but not all) flux ratios. 
We proceed by  
assigning spectral types to the objects found towards the Pleiades, 
and we also derive the strengths of photospheric features, \ha 
and radial velocities. 

\medskip
\centerline{\it 3.1~Pseudocontinuum spectral ratios}

The overall optical-NIR spectrum of very cool stars is largely depressed 
by molecular opacity. The true stellar continuum is never seen, but  
at a few points the molecules are a little more 
transparent and one sees deeper in the photosphere, forming a 
pseudocontinuum (PC hereafter). In our spectral range 
this happens around wavelengths 653.9, 704.5, 756.0, 824.5, 884.0, 920.0 
and 985.0 nm.  

Strong telluric bands (i.e. more than 10\% absorption) 
are present in the following regions: 
685.8--703.2, 715.3--737.7, 757.7--769.2, 809.9--834.5,  
892.0--965.0 and 984.0--1012.0  nm. 
We cancelled out the bulk of the telluric absorption using the flux standard 
HD 19445 (G metal poor star) observed with the same instrumental setting as 
the programme stars. 

Previous works have pointed out that PC ratios are well 
correlated with M spectral subclass 
(Mart\'\i n 1993, Hamilton \& Stauffer 1993). 
We have defined five PC indices, as ratios between the average flux 
at two different spectral regions.   
In Table~2 we specify the integration limits that we have used, and 
in Table~3 we give the values measured in dwarfs of known spectral type 
from  Kirkpatrick, Henry \& Simons (1995). 
All the stars were corrected for their intrinsic doppler velocity prior 
to calculating the ratios.

\begin {table} \caption{Pseudocontinuum Integration Limits (nm)} 
\begin{center} \begin {tabular}{lll}
\hline PC index & Numerator & Denominator \\
\hline
PC1    &  703.0--705.0 & 652.5--655.0 \\
PC2    &  754.0--758.0 & 703.0--705.0 \\
PC3    &  823.5--826.5 & 754.0--758.0 \\
PC4    &  919.0--922.5 & 754.0--758.0 \\
PC5    &  980.0--988.0 & 754.0--758.0 \\
 \hline
\end{tabular} \end{center} 
\end{table}
 
\begin {table} \caption{PC measurements for field dwarfs} 
\begin{center} \begin {tabular}{lllllll}
\hline Name & SpT. & PC1  & PC2 & PC3 & PC4 & PC5 \\
\hline
GJ 873     & M3.5 V     & 1.31  & 1.40 & 1.12 & 1.18 & 1.08 \\
GJ 402     & M4 V       & 1.44  & 1.47 & 1.12 & 1.21 & 1.15 \\
GJ 83.1    & M4.5 V     & 1.40  & 1.50 & 1.13 &      &      \\
GJ 51       & M5 V      & 1.44  & 1.67 & 1.13 &      &      \\
GJ 905      & M5.5 V    & 1.66  & 1.74 & 1.26 & 1.54 & 1.47 \\
GJ 65 AB    & M5.5 V    & 1.56  & 1.91 & 1.29 &      &      \\
LHS 248     & M6.5 V    & 1.82  & 2.10 & 1.45 & 2.12 & 2.22 \\
VB10        & M8 V      & 2.20  & 2.43 & 1.82 & 2.79 & 3.41 \\
LP412-31    & M8 V      & 2.13  & 2.51 & 1.84 & 2.79 & 3.44 \\
LHS 2065    & M9 V      & 1.60  & 1.94 & 1.91 &      &      \\
PC 0025+0447  & M9.5 V  & 1.87: & 1.76: & 2.11 & 3.91 & 6.75: \\
BRI 0021-0214  & M9.5 V & 2.09  & 1.88 & 2.42 & 4.11 & 5.86 \\
\hline
\end{tabular} \end{center} 
\vskip2truept
Notes: The average error bar for the PC indices is \x 0.1, 
except for the values marked with a colon, which have a factor 
of 2 larger uncertainty. 
The PC4 and PC5 indices could not be measured in the WHT spectra because 
they have shorter spectral range than the Calar Alto spectra. 
The spectral types were taken from Kirkpatrick et al. (1995).   
\end{table}

 The PC1 and PC2 ratios have similar behavior:  
they increase with increasingly late spectral type 
until $\sim$ M8 and decrease afterwards. This ``saturation" effect 
has also been noticed in the (R--I) color by Bessell (1991).  
The double-values of PC1 and PC2 make them unreliable for spectral 
type classification, and hence we do not use them for that purpose. 
The other three indices do not present such saturation, as they 
do increase monotonically up to M9.5. Their dependence with spectral type 
can be well fitted by second order polynomials giving the following 
relationships: 

$$ SpT = -8.009 + 14.080 \times PC3 - 2.810 \times PC3^2 ~~N=11~~\sigma=0.39 $$

$$ SpT = -0.944 + 4.663 \times PC4 - 0.515 \times PC4^2 ~~N=8 ~~\sigma=0.22 $$
   
$$ SpT = 1.038 + 2.979 \times PC5 - 0.264 \times PC5^2 ~~N=7 ~~\sigma=0.32 $$
  
where SpT is the M subclass number, N is the number of data points, and 
$\sigma$ is the standard deviation in spectral type subclass 
of the polynomial fit to the PC values.

\medskip
\centerline{\it 3.2~Spectral types for Pleiades objects}

We have used the PC ratios measured in stars of known spectral type 
to calibrate them. Thus, we can apply the relationships given above 
to derive spectral types for new objects. 
Unfortunately, we could not measure the PC4 and PC5 indices  
in our WHT spectra as they do not go far enough into the red. 
The only exception is Roque~1, which was observed at the WHT but the 
spectral range does include the PC4 index. 
We note that for Teide~1, our spectral 
type is solely based on the PC3 index. Rebolo et al. (1995) 
estimated M8.5--M9 from two different calibrations, which is consistent 
within the uncertainties 
with the spectral type derived in this work.  

For those objects observed 
at Calar Alto we took the average M subclass obtained from the 
PC3, PC4 and PC5 values. 
The 1$\sigma$ dispersion was typically 0.3 spectral 
subclasses. Thus, we estimate that the spectral types given in Table~3 
are safe to better than half a spectral subclass. 
In Table~4 we give the PC ratios measured in the Pleiades objects, and 
the spectral types that we infer from them. 
In Figure 3 we plot the PC2 and PC3 indices versus spectral type 
for the programme stars. Different symbols are used to distinguish 
the Pleiades objects from the field dwarfs. We have plotted these 
two ratios to illustrate how one (PC2) is not a good indicator 
of spectral type in the range M5--M10, whereas the other (PC3) is indeed a 
good indicator in the same range.

\begin {table} \caption{PC measurements and spectral types 
for Pleiades objects} 
\begin{center} \begin {tabular}{lllllll}
\hline Name & SpT. & PC1  & PC2 & PC3 & PC4 & PC5 \\
\hline
Calar 7    &  M4       & 1.35  & 1.71 & 1.12 &  &  \\
Calar 4     &  M5       & 1.46  & 1.68 & 1.29 & 1.22 & 1.33: \\
PPl 14      &  M5.5     & 1.71  & 1.95 & 1.27 &  &  \\
Calar 6   &  M5.5     & 1.51  & 2.04 & 1.30 &  &  \\
Calar 2    &  M6       & 1.74: & 1.88: & 1.41 & 1.70 & 2.17:: \\
PPl 15      &  M6.5     & 1.63  & 2.20 & 1.44 &  &  \\
PPl 1       &  M6.5     & 1.82::  & 2.77:: & 1.46: &  &  \\
Calar 5    &  M6.5     & 1.51: & 2.04 & 1.52 & 1.86 & 2.27: \\
Roque 1     &  M7       & 1.50  & 2.21 & 1.60 & 2.38 &  \\
Teide 1     &  M8       & 2.25  & 2.65 & 1.73 &  &  \\
Calar 3    &  M8       & 2.36  & 2.84 & 1.77 & 2.84 & 3.17: \\
Calar 1    &  M9       & 2.84  & 2.21 & 1.99 & 3.31 & 4.20: \\
\hline
\end{tabular} \end{center} 
\vskip2truept
Notes: The average error bar for the PC indices is \x 0.1, 
except for the values marked with one or two colons, which have factors 
of 2 and 3 larger uncertainty, respectively. 
The PC4 and PC5 indices could not be measured in the WHT spectra because 
they have shorter spectral range than the Calar Alto spectra. 
The uncertainty in spectral type is half a spectral subclass.  
\end{table}

\medskip
\centerline{\it 3.3~Absorption Features}

The main molecular bands in the optical-NIR spectrum of our programme 
stars are due to CaH, TiO and VO. 
Kirkpatrick et al. (1991) defined the A-index for measuring 
the strength of CaH at 675--705~nm. In Table~5 the A-index 
measured in our spectra are brought 
together with the VO-index of Kirkpatrick et al. 
(1995) and the TiO-index of Mart\'\i n \& Kun (1996). 

\begin {table} \caption{Absorption Features} 
\begin{center} \begin {tabular}{llllll}
\hline Object & A & TiO & VO & KI  & NaI  \\
 &  &  & & (\AA ) & (\AA ) \\
\hline
GJ 873         & 1.33   & 1.47 & 1.00 & 11.7 & 3.2 \\
GJ 402         & 1.29   & 1.56 & 1.00 & 10.6 & 3.9 \\
GJ83.1         & 1.40   & 1.66 & 1.00 & 11.5 & 6.1 \\    
GJ 51          & 1.41   & 1.77 & 1.00 & 11.9 & 6.0 \\
GJ 905         & 1.32   & 1.81 & 1.02 & 18.1 & 5.7 \\
GJ 65 AB       & 1.50   & 1.92 & 1.02 & 15.6 & 6.6 \\
LHS 248        & 1.57   & 2.15 & 1.05 & 28.3 & 7.4 \\    
VB10           & 1.42   & 1.70 & 1.12 & 18.0 & 6.3 \\ 
LP412-31       & 1.53   & 1.63 & 1.12 & 17.2 & 7.4 \\
LHS 2065       & 1.24   & 1.26 & 1.13 &  9.5 & 4.7 \\
PC 0025+0447   & 1.21   & 0.61 & 1.17 & 10:  & 2:: \\
BRI 0021-0214  & 1.35   & 1.02 & 1.11 & 16.2 & 3.9 \\
Calar 7       & 1.30:  & 1.47:  & 1.02 & 14:: & 5.6: \\
Calar 4        & 1.36:  & 1.67:  & 1.00 & 11:  & 3.5:: \\
PPl 14         & 1.33   & 2.03   & 1.04 & 22.8: & 5.8 \\ 
Calar 6      & 1.40:  & 1.77:  & 1.01 & 21.3: & 6.5 \\
Calar 2       & 1.36:: & 2.99:: & 1.04 & 20.2: & 7.2 \\
PPl 15         & 1.51   & 2.26   & 1.06 & 21.0: & 4.8 \\ 
PPl 1          & 1.14::   & 2.47::   & 1.01:: &  20:: & 8.8:: \\
Calar 5       & 2.98:: & 1.63:  & 1.05 & 21.3: & 5.2: \\
Roque 1        & 1.89   & 2.50: & 1.03 & 23.5: & 8.9 \\
Teide 1        & 1.48   & 1.55   & 1.15 & 16.8: & 5.1 \\
Calar 3       & 1.20:  & 1.58:  & 1.16 & 20.2: & 4.5: \\
Calar 1       & 1.40   & 1.44   & 1.07 & 30::  & 6.6 \\
\hline
\end{tabular} \end{center} 
\vskip2truept
Notes: The error bars are 
\x 0.5 and 1 \AA ~for the NaI and KI EWs, respectively, and 
5\% for the indices, except for the values marked 
with one or two colons, which have factors of 2 or 3 larger uncertainty. 
\end{table}

The strongest atomic lines in our spectral range are the NaI doublet 
at 818.3 and 819.5 nm, and the KI resonance doublet at 766.5 and 769.9 nm. 
In Table~5 we give the equivalent widths (EW) of these atomic lines, which 
were obtained by direct integration of the two lines (partially blended 
in our spectra). The continuum levels for the EWs of the 
KI and NaI doublets were set at the average flux in 
30~\AA ~around 764.5 and 814.0~nm, respectively. 
Note that these are not the true continuum for the absorption lines, 
but the observed pseudocontinuum. Our EW values should be considered as 
lower limits to the true equivalent widths. 
 
 The NaI doublet and the KI line at 766.5 nm are affected 
by telluric bands which absorb up to 15\% of the continuum.   
We have estimated that if we had not corrected from 
telluric absorption our EWs of the NaI doublet 
would be larger by $\sim$10\% . Note for instance that the C-index 
of Kirkpatrick et al. (1991), which measures the NaI doublet,  
was not corrected for telluric absorption, and 
their published values should be decreased by about 10\% . 
We find a good agreement for the stars in common after allowing 
for this correction. The effect of telluric absorption on the KI doublet is 
more dramatic, and we find in general that the EWs are a factor of 
$\sim$2 larger after correction. 
The NaI doublet in some spectra from Calar Alto  
are affected by one bad pixel in the detector. We substituted the bad 
pixel by the interpolated value of the two adjacent ones.

\medskip
\centerline{\it 3.4~ \ha Emission and Radial Velocities}

 We derived  
the \ha EW in our spectra via direct integration of the line profile. 
The results are given in Table~6. We give the measurements of 
individual exposures when \ha was variable. For Teide~1 
the average \ha EW of seven spectra is 6 \AA  ~(Rebolo et al. 1995). 
In Table~6 we give the \ha EWs for the three best individual spectra 
(each 1 hour integration time at the WHT). We have noted significant  
\ha variability in LHS~248, VB10, PC 0025+0447 and Teide 1.  

Radial velocities were computed via cross-correlation of the 
spectra with templates of similar spectral type. We used as 
templates LHS 248 (Basri \& Marcy 1995), VB10 (Goldberg 1995),  
and BRI 0021-0214 (Basri \& Marcy 1995). For LHS 2065 
and Teide 1 we adopted the radial velocities given by 
Mart\'\i n, Rebolo \& Magazz\`u (1994) and Rebolo et al. (1995), respectively.  
The radial velocities obtained for the programme stars 
and the reference star used are given in Table~6. 
The spectral window selected for the cross correlation was 840--880 nm, 
which is free for telluric bands and contains a lot of photospheric lines   
in late-type spectra (Mazeh et al. 1996).

\begin {table} \caption{\ha Emission and Radial Velocities} 
\begin{center} \begin {tabular}{llccl}
\hline 
Object & JD & \ha  & Vrad & Template \\
       & --2400000 & (\AA ) &  (\kms ) \\
\hline
GJ 65 AB      & 50020.78 & 8.2 & 62 & LHS 248 \\
LHS 248       & 50020.65 & 3.3 & 9$^*$  &         \\
              & 50020.77 & 5.5 &   &         \\
VB10          & 50019.35 & 2.9 & 35$^*$ &         \\
              & 50019.37 & 6.3 &  &         \\    
LP412-31      & 50018.68 & 28.6 & 1 & VB10    \\
LHS 2065      & 49716.56 & 10.7 & 4.5$^*$ &         \\
PC 0025+0447  & 50020.38 & 475 & 46 & BRI 0021-0214 \\
              & 50020.40 & 366 & 20 & BRI 0021-0214 \\
BRI 0021-0214 & 50018.06 & $<$0.4 &  13$^*$ & \\
              & 50019.49 & $<$0.1 &   & \\
Calar 7      & 50020.70 & $<$0.5 & -109: & LHS 248 \\
Calar 4       & 50020.56 & $<$2.5 &  &  \\
PPl 14        & 50019.52 & 3.7 & 22: & LHS 248 \\
Calar 6     & 50020.70 & $<$0.5 & -62 & LHS 248 \\
Calar 2      & 50020.56 & $<$1.5 & -5: & LHS 248 \\
PPl 15        & 50020.61 & 11.5: & 20 & LHS 248 \\
PPl 1        & 48645.43 &  6.5: & 14$^*$ &  \\
Calar 5      & 50020.61 &  6.5 &  -8: & LHS 248 \\
Roque 1       & 50131.49 &  $\le$1.5 & 151  & LHS 248 \\
Teide 1       & 49716.41 &  3.5: & 2$^*$  &  \\
              & 49716.46 &  3.7: &  &   \\
              & 49716.50 &  8.6 &   &   \\
Calar 3      & 50020.51 & 10.2 & -14 & VB10 \\
Calar 1      & 50019.65 & $\le$2.5 & 85 & VB10 \\
\hline
\end{tabular} \end{center} 
\vskip2truept
Notes: The average error bars are  \x 1.0 \AA ~ for the equivalent widths, and 
30 \kms for the radial velocities, except where marked 
with colons, which have error bars a factor of 2 larger. 
The asterisks denote radial velocities 
taken from the literature (see text). The spectra of 
Calar 4 and PPl 1 were too noisy for deriving a radial velocity.  
\end{table}

\medskip

The velocity dispersion of our spectra is rather poor for radial velocity 
work (137.5 \kms pix$^{-1}$ at Calar, and 116 \kms pix$^{-1}$ 
at the WHT). Nevertheless, the cross correlation technique was able to achieve 
precisions of about 1/4 of a pixel, so we obtained radial velocities 
accurate to about 30 \kms . We verified this by cross correlating 
the standards against each other. The main limitation on the 
radial velocity accuracy comes from the broadness of the cross-correlation 
peak due to the shape of the molecular absorption bandheads. 
It is also possible that part of the error comes from mismatches between 
the target and the template. We correlated every target  with its closest 
proxy in spectral type.    
For Calar 4 and PPl~1 
the cross-correlation peaks 
were too broad and asymmetric and we could not derive radial velocities 
for them. The radial velocity of PPl~1 has been measured by Stauffer et al. 
(1994b), and that is the value quoted in Table~6.

\bigskip 

\centerline{\large 4.~DISCUSSION}

\medskip
\centerline{\it 4.1~Cluster Membership}

The key question about the nature of our eight new BD candidates is whether 
they belong to the Pleiades. Photometrically they occupy 
the region of the (R--I) vs I diagram where substellar cluster members 
are expected to be located (Zapatero-Osorio et al. 1996). However, 
this is not a strong enough constraint as other types of objects could 
have similar photometric properties (reddened galaxies, background 
red giants and late-type dwarfs, foreground very late-type dwarfs). 
In this section we use the analysis of the spectroscopic data 
described above to asess the membership status of each Calar object and 
Roque 1. 
We compare them with bona fide very low mass cluster members, and 
in particular PPl 15 and Teide 1. 
Our approach has been ``cautious" in the sense that 
we retained as members only the objects that fully meet 
all the criteria (gravity, radial velocity, \ha, spectral sequence). 
Thus, we may have left out possible members rather than to 
accept field stars as members. 

First, we look at the gravities. 
An interesting piece of information is the fact that all the Pleiades BD 
candidates present KI and NaI photospheric features (Section 3.2), 
implying that they are dwarfs (cf. Kirkpatrick et al. 1991). None of them 
is a red giant or an extragalactic object. This result indicates that 
the region of BDs in 
the I-band vs. (R--I)-color diagramme of the Pleiades cluster is not 
significantly contaminated by giants and extragalactic objects. 

A kinematic constraint on membership comes from the fact that 
isolated brown dwarfs 
are thought to be formed as independent condensations in a manner similar 
to low mass stars, and 
hence they should also share the bulk motion of the cluster. 
The radial velocities of known members of the Pleiades are in the 
range 0--14 \kms ~(Stauffer et al. 1994b).  Taking into account the 
radial velocities given in Table~6 and their error bars, we consider 
as possible kinematical members Calar 2, 3 and 5. The other objects  
could be field stars, or large-amplitude spectroscopic binaries, or even 
run-away objects.  However, from a ``cautious" 
point of view we just consider them as kinematical non-members 
in the third column of Table 7.  

Another criterion of membership in the Pleiades is the presence of \ha 
in emission. Hodgkin, Jameson \& Steele (1996) 
measured the \ha strength of 6 proper motion 
members with I magnitude $\ge$ 16.5. All of them had \ha emission 
with equivalent widths in the range 3.1--6.8~\AA . Stauffer et al. (1994a,b) 
found a similar range of \ha equivalent widths among their Pleiades PPl stars. 
We find that PPl~1, PPl~14, PPl~15 and Teide 1 
show similarly strong \ha lines (Table~6). 
However, among field very late type stars, there are 
many cases of weak \ha emission. For instance Mart\'\i n et al. 
(1994) found \ha EW less than 3 \AA ~for 5 out of 10 M7--M9 field dwarfs.  
To summarize, we consider that \ha emission equivalent width $<$~3 \AA 
~ indicates that an object of spectral type in the range M4--M8 is 
probably not a Pleiades member. The value of 3 \AA ~ is somewhat arbitrary, 
but it reflects the lower envelope of the \ha equivalent widths 
measured among Pleiades very low mass objects. We have to make an exception 
with Calar~1 as it has the latest 
spectral type known in a dwarf towards the Pleiades cluster. We do not 
know what to expect for \ha in objects cooler than Teide~1, and therefore 
this criterion cannot be applied for Calar~1. 

Usually cluster members are expected to follow a tight 
sequence in the H-R diagram. PPl~15 and 
Teide 1 extend the Pleiades sequence towards magnitudes as faint as 
the new BD candidates. 
This is illustrated in Figure~4 where we display  
apparent I-magnitudes versus spectral types. 
We have plotted  the proper motion members (HHJ~stars) with spectral types 
later than M4  (determined by Mart\'\i n et al. 1994 for HHJ~10, and 
Steele \& Jameson 1995 for the rest), three PPl objects and Teide 1. 
A spectral sequence is clearly defined by these objects. 
We have also plotted the new BD candidates using different symbols 
for clarity. Only Calar~3 and Roque~1 nicely fit with the  
spectral sequence. Calar~3 is indeed very close to Teide 1. 
The rest of the Calar objects lie below the sequence, except Calar~1. 
We note that one of the HHJ stars is also well below the sequence. 

There is some controversy about a possible age 
spread in the Pleiades cluster (e.g. Stauffer et al. 1994). 
This issue is not yet settled, and in fact 
it could partially explain the spread seen in Figure~4. 
Very low mass Pleiades objects are gravitationally contracting 
and hence their luminosity and temperature are time-dependent. 
We have used a grid of evolutionary tracks kindly provided by 
F. D'Antona with a time step of 10 Myr in the mass range 
0.1--0.06~M$_\odot$ to estimate the change in I-band magnitude 
due to a possible range of ages of 70--130~Myr among Pleiades objects. 
We have obtained a maximum difference in I-band of $\sim$ 0.7 mag 
(shown in Figure~4 as a vertical dotted line). In fact, among the 
M5.5 and M6.5 Pleiades members the spread in I-band magnitudes 
seems similar to that  
expected for a 60 Myr age spread. Such an age spread may not be real, 
but so far it cannot be ruled out.  
BD candidates like Calar~1, 2 and 5 could be consistent with 
the spectral sequence if there is such an age spread and taking into account 
the error bars of $\sim$ 0.1 mag in I-band and half a spectral subclass 
in spectral type. Moreover, Zapatero-Osorio et al. (1996) have shown 
that Calar~2 and 5 are affected by abnormal reddening probably due 
to a small interstellar cloud. The de-reddening correction is about 
0.5 magnitudes in the I-band. Thus, we cannot rule out that these objects 
are cluster members just from Fig.~4. 

\begin {table} \caption{Membership Status} 
\begin{center} \begin {tabular}{llllll}
\hline 
Object    & Gravity & Vrad & \ha & Sp.Seq. & Final \\
\hline
Calar 1    & Yes     & No   & ?  & Yes? & No \\
Calar 2    & Yes     & Yes  & No  & No? & No? \\
Calar 3   & Yes     & Yes  & Yes & Yes & Yes \\
Calar 4     & Yes     & ?    & No  & No & No \\
Calar 5   & Yes     & Yes  & Yes & No?  & No? \\
Calar 6  & Yes     & No   & No  & No  & No \\
Calar 7   & Yes     & No   & No  & No  & No \\
Roque 1   & Yes     & No   & No  & Yes  & No \\
\hline
\end{tabular} \end{center} 
\end{table}

Our ``cautious" conclusions regarding membership for the objects  
are given in the sixth column 
of Table 7. We see that there are some clear-cut cases where all  
membership criteria agree. For instance Calar 6 and 7 have spectral 
types and radial velocities inconsistent with membership, and moreover 
\ha emission is not detected, so they are obvious non-members. On the other 
hand, Calar~3 is the only object for which all the criteria are 
affirmative so it is very likely a cluster member. 
However, there are also ambigous cases where 
the criteria disagree. We discuss them individually: 

{\bf Calar 1:} It is the coolest dwarf found in our survey (M9). However, 
it appears to be too bright for its spectral type in Figure~4, and 
it has too high a radial velocity. If it is a binary with almost 
equal mass components, it would be overluminous and if the orbital period 
is short its radial velocity  would be variable. 
However, our spectra do not show any hint for radial velocity variability, 
nor the presence of line splitting.   
Tentatively, we consider this object as a non-member for further discussion 
in this paper, but it is certainly worth to check if it is a spectroscopic 
binary. Zapaterio-Osorio et al. (1996) estimated that the probability of 
finding an M9 field dwarf in their survey is 
only $\sim$ 10\% on the basis of the density of M9 dwarfs in the solar 
neighbourhood. However, we note that the high radial velocity of Calar 1 
is more typical of an old disk star than of a young disk star. 
It may be that the contribution of old disk M9 dwarfs starts to become 
significant in a survey like that of Zapatero-Osorio et al. which goes 
much deeper than larger-area surveys in the solar neighbourhood.  
Alternatively, Calar~1 may have been born in an unstable 
triple system, and has been ejected from it by interaction with more 
massive companions.  
In any case, this object is one of the few M9 dwarfs known in the whole 
sky, and deserves further investigation.  

{\bf Calar 2:} A dwarf with radial velocity consistent with membership. 
However, given our large error bar on the radial velocity this is a weak 
constraint. We give more weight to the fact that lies out 
of the spectral sequence of Figure 4 (although marginally consistent 
within the uncertainties), and it does not show \ha in emission. 
Thus, we consider that this object is more likely a background old dwarf 
than a cluster member, but we cannot reach a definitive conclusion with 
our data. 

{\bf Calar 5:} Its gravity, radial velocity and \ha are all supportive 
of membership, 
but it does not fit the spectral sequence. 
As mentioned before, Calar~5 is affected 
by larger reddening than the average in the Pleiades cluster, probably 
due to a small interstellar cloud (Zapatero-Osorio et al. 1996). 
However, the de-reddening correction is not enough to put it 
in the region of Pleiades members in Figure~4. This casts serious doubts 
on its membership.  
Together with Calar~2, Calar~5 would need more data  
(proper motion, accurate radial velocity) in order to finally 
establish whether or not it belongs to the cluster.  
Interestingly, there is one proper motion member 
(HHJ~7) which lies in the same region of Figure~4 as Calar~2 and 5. 
HHJ~7 has weak \ha emission (Hodgkin et al. 1996) and according to our 
membership criteria we would consider it as a probable cluster non-member. 
However, if 
future observations establish that HHJ~7, Calar~2 and 5 have radial 
velocities and proper motions consistent with membership,  
the important implication would be that there is a very large age spread 
among the Pleiades very low mass objects, or else there is a background 
population that has a similar space motion. 

{\bf Roque 1:} Despite of being an M7 dwarf which nicely fits in the spectral 
sequence for the Pleiades cluster, we believe it is probably 
not a cluster member 
because it does not show \ha emission and 
has a much too high radial velocity. To our knowledge, its radial velocity 
is the highest ever measured in an M7 dwarf, and may indicate 
membership to an old galactic population. However, we do not find in the 
spectrum any clear indication of a low metallicity.

\medskip
\centerline{\it 4.2~Spectroscopic Properties of Young Brown Dwarfs}

The study of very low mass stars 
and brown dwarfs is complicated mainly because of the formation of 
many molecules and possibly dust in their atmospheres,  
which shape the observed spectrum. Theoretical models have tried to 
reproduce the spectral distribution of well-known very cool dwarfs like VB10, 
with not completely successful results, e.g. Allard \& Hauschildt (1995).  
In consequence, 
there is not a consensus on how to derive basic parameters like 
effective temperature, gravity and metallicity. From the study of 
low-luminosity objects in clusters of known age, distance and metallicity, 
we hope to learn something about the structure and evolution of 
their atmospheres. In this work we can compare the spectra of the 
Pleiades very low mass objects with some of the very late-type field dwarfs. 

In the previous section we have concluded that one of the BD 
candidates is very likely a twin of Teide~1. These two objects 
define the cool end of the presently known spectral sequence for 
Pleiades members. 
Recent applications of the Li-test have shown that this sequence 
goes beyond the substellar limit. 
Lithium has been detected in PPl 15 (Basri et al. 1996) 
and Teide 1 (Rebolo et al. 1996), but not in HHJ3 and HHJ10 
(Marcy, Basri \& Graham 1994; Mart\'\i n et al. 1994),  
which have spectral types of M6 (Steele \& 
Jameson 1995) and M5.5 (Mart\'\i n et al. 1994), respectively. 
 The lack of lithium implies that Pleiades objects 
earlier than $\sim$ M6.5 are probably stars. 
The Li detection in PPl~15, although 
with a depleted abundance, indicates that the transition between 
stars and brown dwarfs takes place at spectral types around M6.5. 
Pleiades members with spectral types M7 and later are very likely brown dwarfs, 
as confirmed by the Li detection in Teide~1. 
  
We can learn about the atmospheres of young brown dwarfs by 
comparing them with field stars, and vice versa. In general the 
spectroscopic properties seem dominated by the effective temperature, 
because young brown dwarfs and M8--M9 field dwarfs, 
have similar overall spectral characteristics. 
 However, a close look reveals that there are some differences:   

\noindent
{\bf $\bullet$} 
The Pleiades very low mass objects tend to have lower values of NaI EW than 
the M6--M9 standards  (Figure 5a). This trend was 
also noted by Steele \& Jameson (1995) among M2--M6.5 Pleiades stars. 
The NaI doublet is known to be gravity-sensitive because it is 
very weak in giants. Since Pleiades very low mass stars and 
BDs have lower gravities than 
main sequence stars by about a factor of only $\sim$ 2, 
the observed weakening of the NaI lines confirms that they are quite 
sensitive to small changes in gravity. 

\noindent
{\bf $\bullet$}
 The VO index tends to be slightly higher in the Pleiades objects than 
in the field dwarfs (Figure 5b), but not the A (CaH) and TiO indices. 
We interpret this result as follows: the CaH and TiO molecular bands 
have similar strengths because the standards have the same 
metallicity as the Pleiades, i.e. solar, but the VO-index is different 
because it may have some sensitivity to gravity. An alternative 
interpretation of the VO difference between Pleiades and field dwarfs 
has been suggested by the referee of this paper. It may be that 
the indices that we have used to compute the spectral types 
(PC3, PC4, PC5) carry systematic errors when applied to low gravity 
Pleiades objects. If that were the case, the VO-index would suggest 
that Teide~1 and Calar~3 are about 1 spectral subclass later, i.e. 
M9 instead of M8. Such a consideration warns one of the limitations of 
using spectral types which are originally defined for field dwarfs 
for another class of objects as young brown dwarfs.   

\noindent
{\bf $\bullet$}
 The pseudocontinuum in the range 650--750 nm is more depressed in Teide 1 
and Calar~3 than in the M8 standards, 
thereby explaining their also higher 
(R--I) color (Zapatero-Osorio et al. 1996).
Possible reasons for these color differences are gravity-dependent 
formation of dust in cool photospheres (Tsuji, Ohnaka \& Aoki 1996), and/or   
the presence of circumstellar material. 
The second hypothesis 
would also help to explain the large spread in I-magnitudes 
for a given spectral type (Figure~4), without invoking a large age spread. 
It will be very interesting to test the presence of dust and 
circumstellar disks with IR photometry and spectroscopy. 

\medskip
\centerline{\it 4.3~Future CCD Brown Dwarf Surveys}

The results obtained in this work could be used to improve future 
photometric searches. The success rate of the broad-band CCD 
survey of Zapatero-Osorio et al. (1996) is $\sim$25\%. 
Other photometric 
surveys do not seem to have a higher success rate. 
For instance, Mart\'\i n 
(1993) obtained optical spectra for five BD candidates from the (I--K) 
survey of Simons \& Becklin (1992) and found no dwarf later than M4  
among them.  Hamilton \& Stauffer (1993) obtained optical spectra for 
seven objects with I=16--17.8, and found only one with spectral type 
later than M6 (PPl~1).  

Proper motion surveys have a better chance of success (Hambly, 
Hawkins \& Jameson 1993; Rebolo et al. 1995),  
but they need images at two epochs separated 
by many years because the mean proper motion of the Pleiades is only 
0.05 arcsec yr$^{-1}$. Steele \& Jameson (1995) obtained 
low-resolution spectroscopy of 33 proper motion members with I=13.5--17.5, 
and classified them as dwarfs with spectral types in the range M2--M6.5 
from comparison with a few reference stars.  

Based on the shape of the spectra that we have analyzed in this work,  
we propose that CCD photometric surveys could use filters 
with moderately narrow passbands. Taking into account the pseudocontinuum 
points of the M8--M9.5 dwarfs, and trying to avoid regions of strong 
telluric absorption and sky airglow we have selected the following 
passbands: 730--758~nm, 810--838~nm and 900--928~nm. 
Hence, the filters that we 
propose are 28~nm wide only, which is a factor of $\sim$ 8 less than  
standard R,I filters. There is a loss of efficiency but 
it is not as large as it may seem, because we have chosen  
spectral regions of maximum emission in the brown dwarfs. 
As a further compensation we may add that the use of our 
filters would allow one to carry out the CCD surveys in non-dark 
conditions (moon background, light pollution), which are needed   
 if using for instance the R-band. 
Moreover, two problems of using broad-band filters in large aperture 
telescopes would be alleviated; the saturation of bright stars and 
the high level of sky background in the near-IR.    
 
While the (R--I) color is known to saturate for spectral types later 
than $\sim$ M7 (Bessell 1991), 
simulations (using spectra) of the colors given by our filters  
indicate that they increase motonically from M4 to M9.5, and thus it 
would be useful to discriminate between BDs and background mid-M dwarfs 
in young open clusters. 
The filters would 
have to be specially manufactured, but, since follow-up 
 spectroscopy of very faint BD candidates consumes a lot of 
large-aperture telescope time, we believe that future photometric 
searches should be optimized by using narrow pseudocontinuum filters 
like the ones suggested by us. 

Another possible way of increasing 
the success of photometric surveys is to combine optical, near-IR and 
IR filters, such as for instance standard broad-band R, I, J and K filters.  
The disadvantage would be that multi-band observations require more 
telescope time and calibration efforts. 

\bigskip 

\centerline{\large 5.~CONCLUSIONS}

We have obtained optical-near-IR spectroscopy of eight new photometric 
BD candidates in the Pleiades cluster found by  
Zapatero-Osorio et al. (1996). In addition we observed  
the Pleiades objects PPl~1, 14, 15, and Teide~1, and several well-known 
very late-type standards. We have used three relationships between 
pseudocontinuum ratios and spectral type, in order to classify the 
Pleiades objects. We have derived the absorption strength 
of molecular (CaH, TiO, VO) bands and the equivalent widths of 
photospheric atomic features (KI, NaI), and chromospheric \ha emission. 
We have also obtained radial velocities although  with considerable 
uncertainty (typically \x 30 \kms ) due to our low spectral resolution.  

One of the new BD candidates, namely Calar~3, has a similar I-band  
magnitude and spectral type as Teide 1 (M8). It also has \ha emission 
and radial velocity consistent with membership in the Pleiades. 
Hence, this object is very likely a twin of Teide 1, 
i.e. a new Pleiades brown dwarf. 
Other interesting objects are Calar~1, 2, 5 and Roque~1. 
Calar~2 and 5 might 
be cluster members (consistent radial velocities) but they are  
about 1 magnitude too faint in the I-band for their spectral types 
(M6, M6.5), and a rather large age spread would have to be invoked 
if they were members. 
Calar~1 is the coolest dwarf ever found towards 
the Pleiades (M9). However, its radial velocity is not at all consistent 
with membership.  It is remarkable that Roque~1 has an even higher velocity. 
The presence of two very late dwarfs with high velocity was unexpected 
in the small volume surveyed by Zapatero-Osorio et al.  
The rest of the BD candidates have too early spectral types (M4--M6) 
and in some cases too high radial velocities, and we believe that they are 
background stars which have contaminated the photometric survey.    

PPl~15 (M6.5), Teide~1 (M8) and Calar~3 (M8) define 
a spectral sequence at the substellar mass limit and beyond in the Pleiades 
cluster, 
according to the lithium test. Comparison of their spectra with 
field dwarfs indicate that in general they are very similar,but we have 
tentatively identified a few subtle differences:  
 In the Pleiades BDs, the NaI doublet tends to be weaker 
and the VO-index stronger than in the field stars. 
These differences are probably due to lower gravity in the Pleiades objects, 
as they are young and still contracting. The difference in VO also 
warns us of the limitations of using the spectral type classification 
of field dwarfs for young brown dwarfs. 
We have also remarked on a difference 
in the pseudocontinuum in the region 650--750~nm, in the sense that 
the M8 Pleiades tend to have higher PC2 indices. We suggest that it could 
be due to gravity-sensitive dust formation, and/or circumstellar extinction. 

Our results indicate that background 
late-type dwarfs (M4--M6) are a significant source of contamination 
($\sim$50\% )
in deep CCD-based surveys using broad-band filters. We argue  
that narrow-band indices centered at pseudocontinuum points, 
especially redwards of 700~nm, allow one 
to discriminate between mid-M contaminants and very late-M dwarfs. 
Hence, we 
propose the use of special filters for improving the success rate of 
future CCD brown dwarf searches in young open clusters.  

\bigskip

\centerline{\large ACKNOWLEDGEMENTS}

We thank Gibor Basri, Franca D'Antona and Dorit Goldberg for sending us 
results prior to publication. 
This work has been partially supported by the Spanish DGICYT project No.  PB92-0434-C02. 

\newpage

\section*{References} \begin{trivlist}

\item [] Allard, F. \& Hauschildt, H.P. 1995, in 
The Bottom of the Main Sequence-- and Beyond, ed C.G. Tinney, Springer, p. 32 

\item [] Basri, G. \& Marcy, G.W., 1995, AJ, 109, 762

\item [] Basri, G., Marcy, G.W. \& Graham, J.R. 1996, ApJ, 458, 600

\item [] Bessell, M.S. 1991, AJ, 101, 662

\item [] Goldberg, D. 1995, private communication

\item [] Hambly, N.C., Hawkins, M.R.S.  \&  Jameson, R.F. 1993, A\&AS, 100, 607

\item [] Hamilton, D. \& Stauffer, J.R., 1993, AJ, 105, 1855

\item [] Hodkin, S.T., Jameson, R.F. \& Steele, I.A. 1996, MNRAS, in press

\item [] Jameson, R.F. 1995, in 
The Bottom of the Main Sequence-- and Beyond, ed C.G. Tinney, Springer, p. 183 

\item [] Kirkpatrick, J.D, Henry, T.J. \& McCarthy, Jr., D.W. 1991, ApJS, 77, 417 

\item [] Kirkpatrick, J.D, Henry, T.J. \& Simons, D.A. 1995, AJ, 109, 797 

\item [] Leggett, S.K. 1992, ApJS, 82, 351 

\item [] Marcy, G.W., Basri, G., \& Graham, J.R. 1994, ApJ, 428, L57 

\item [] Mart\'\i n, E.L. 1993, PhD Thesis, Universidad de La Laguna 

\item [] Mart\'\i n, E.L. \& Kun, M. 1996, A\&AS, 116, 1

\item [] Mart\'\i n, E.L., Rebolo, R. \& Magazz\`u, A. 1994, ApJ, 436, 262 

\item [] Mazeh, T., Mart\'\i n, E.L., Goldberg, D. \& Smith, H. 1996, MNRAS, 
in press 

\item [] Rebolo, R., Magazz\`u, A. \& Mart\'\i n, E.L.   1995, in 
The Bottom of the Main Sequence-- and Beyond, ed C.G. Tinney, Springer, p. 159 

\item [] Rebolo, R., Mart\'\i n, E.L.  \& Magazz\`u, A.  1992, ApJ, 389, L83 

\item [] Rebolo, R., Zapatero-Osorio, M.R. \&  Mart\'\i n, E.L.  1995, Nature, 377, 129 

\item [] Rebolo, R., Basri, G., Marcy, G., Mart\'\i n, E.L., Zapatero-Osorio, 
M.R. 1996, in preparation 

\item [] Simons, D.A.  \& Becklin, E.E. 1992, ApJ, 390, 431

\item [] Stauffer, J., Hamilton, D. \& Probst, R. 1994a, AJ, 108, 155

\item [] Stauffer, J., Liebert, J.R., Giampapa, M., Macintosh, B., Reid, N. 
 \& Hamilton, D. 1994b, AJ, 108, 160

\item [] Steele, I.A.  \& Jameson, R.F. 1995, MNRAS, 272, 630

\item [] Tsuji, T., Ohnaka, K. \& Aoki, W. 1996, A\&A, 305, L1

\item [] Zapatero-Osorio, M.R., Rebolo, R. \& Mart\'\i n, E.L. 1996, A\&A, 
in press

\end{trivlist}

\newpage

\centerline{\large Figure Captions:}

{\bf Figure 1.} Final spectra of Pleiades and field  
very cool dwarfs observed at the WHT with 
ISIS (FWHM=5.8 \AA ). They are ordered by increasingly late 
spectral type from top to bottom. The strongest atomic and molecular 
features are marked. 

\medskip

{\bf Figure 2.} Same as Figure~1 but for targets observed at the 
3.5~m Calar Alto telescope with the TWIN spectrograph (FWHM=7.8 \AA ). 
In some spectra, bad pixels produce spurious features at 798.5~nm, and 
at 818.5~nm. The spectral ranges of the pseudocontinuum ratios 
defined in this work are joined with dotted lines. The actual 
integration regions  
are the narrow solid segments at the edges of each dotted line.  

\medskip

{\bf Figure 3.} The PC2 and PC3 spectral index values measured 
in the field dwarfs (filled hexagons) and in the 
Pleiades objects (open hexagons). Note the different behaviour:  
the PC2 index peaks at around M8 and decreases for later subclasses, 
whereas the PC3 index increases monotonically from M5 to M9.5.  

\medskip

{\bf Figure 4.} I-band apparent magnitudes versus  
spectral type subclass. 
Filled pentagons are used for Pleiades proper motion members from Hambly et al. 
(1993) with spectral types $\ge$ M4 derived by Mart\'\i n et al. (1994)  
and Steele \& Jameson (1995). 
Filled hexagons denote the PPl objects from Stauffer et al. (1994a,b) 
and Teide 1 (Rebolo et al. 1995), with spectral types derived in this work.  
Empty squares denote the BD candidates found by Zapatero-Osorio et al. (1996). 
The names of all objects with spectral type later than M5.5 are indicated. 
In the upper right corner, 
the dotted horizontal line represents the typical uncertainty in 
spectral type and the dotted vertical line the range in I magnitudes 
induced by a hypothetical age spread of 60 Myr. 
For this estimate we used a central age of 100 Myr for the Pleiades and the 
grid of models provided by F. D'Antona.  
The dashed line joins the I-band magnitudes of main sequence 
stars at the distance of the Pleiades, and hence separates 
the region of plausible cluster members (upper part) from that of 
likely background objects (lower part). 

\medskip

{\bf Figure 5.} The NaI equivalent widths (upper panel) and VO-index 
values (lower panel) measured in the standards (filled hexagons) 
and in the Pleiades members PPl~14, PPl~15, Teide~1 and Calar~3 
(open hexagons) plotted against spectral type.

\end{document}